\documentclass[acmsmall]{acmart}

\AtBeginDocument{%
  }

\setcopyright{cc}
\setcctype{by-nc-nd}
\acmJournal{PACMHCI}
\acmYear{2026} \acmVolume{xx} \acmNumber{x} \acmArticle{CSCWxxx} \acmMonth{11} \acmPrice{}\acmDOI{xx.xxxx/xxxxxx}

\usepackage{CJKutf8}
\usepackage{array}
\newcolumntype{L}[1]{>{\raggedright\let\newline\\\arraybackslash\hspace{0pt}}m{#1}}
\newcolumntype{C}[1]{>{\centering\let\newline\\\arraybackslash\hspace{0pt}}m{#1}}
\newcolumntype{R}[1]{>{\raggedleft\let\newline\\\arraybackslash\hspace{0pt}}m{#1}}
\renewcommand{\arraystretch}{1.1}
\usepackage{colortbl} 
\usepackage{adjustbox}
\usepackage{multirow}
\usepackage{svg}
\definecolor{mypink2}{RGB}{239, 240, 255}
\begin{document}

\title[Understanding Player Values and Expectancy for Reporting Systems in Video Games]{``I Don't Have Faith in the Developers to Use My Feedback'': Understanding Player Values and Expectancy for Reporting Systems in Video Games}

\author{Michael Yin}
\affiliation{
  \institution{University of British Columbia}
  \city{Vancouver}
  \state{BC}
  \country{Canada}
}
\email{jiyin@cs.ubc.ca}
\orcid{0000-0003-1164-5229}

\author{Chenxinran Shen}
\affiliation{
  \institution{University of British Columbia}
  \city{Vancouver}
  \state{BC}
  \country{Canada}
}
\email{elise.shen007@gmail.com}
\orcid{0009-0009-9967-8017}

\author{Robert Xiao}
\affiliation{
  \institution{University of British Columbia}
  \city{Vancouver}
  \state{BC}
  \country{Canada} 
}
\email{brx@cs.ubc.ca}
\orcid{0000-0003-4306-8825}

\begin{abstract}
Reporting systems in multiplayer video games allow players to express their dissatisfaction with others and combat in-game toxicity. In this work, we examined the act of reporting through the lens of expectancy-value theory. Using a distributed survey (n = 98) and follow-up interviews (n = 19), we explored the value players place on reporting, their desired outcomes, and their expectations that these outcomes will be achieved. Our findings revealed that reporting is motivated by both altruistic and retributive factors, with players seeking short-term revenge while also looking to foster an improved long-term community. Yet, players felt that reporting may not always meet these goals, with belief in the system being mediated by factors such as developer reputation, reporting transparency, and alignment with the community. By understanding the value and expectancy of reporting systems, we discuss their implications on broader digital moderation and consider current and potential future designs of reporting systems.
\end{abstract}



\received{20 February 2007}
\received[revised]{12 March 2009}
\received[accepted]{5 June 2009}

\maketitle

\section{Introduction}

\textbf{Understanding and handling toxic encounters} within games has long been a challenging area of ludological research. Toxic encounters can manifest in communicative forms, such as insults in voice or text chat, or non-communicative forms, such as intentionally ruining a game through purposefully throwing or quitting prematurely \cite{kouToxicBehaviorsTeamBased2020, holmes2024toxicityinLeague, laatoTraumatizingJustAnnoying2024}. Toxicity is important for game developers to manage, as it can ruin players' experiences as well as propagate negativity, frustration, and vitriol \cite{kordyakaCuringToxicityDeveloping2021, wijkstraHelpMyGame2023, frommelToxicityOnlineGames2024}. Yet within certain gaming communities, toxic behaviour has unfortunately become increasingly normalized, with players viewing it as inherent to a competitive mindset \cite{beresDontYouKnow2021, turkaySeeNoEvil2020}. The primary feature offered to players to deal with toxic encounters is through in-game reporting (or flagging) systems \cite{kouFlagFlaggabilityAutomated2021a, pohjanenReportPleaseSurvey2018}. After a player submits a report regarding a perceived toxic player, it is sent to review --- this system ends up deciding whether a punishment is warranted, and if so, how severe it should be. Although the goal of reporting systems is to combat toxicity, their effectiveness and practical usage remain lingering questions \cite{kouFlagFlaggabilityAutomated2021a, pohjanenReportPleaseSurvey2018, lapollaTacklingToxicityIdentifying2020, hilbersEvaluationEffectivenessSmites2015}. 

For the player who initiated the report, the extent of how they can combat a toxic encounter within a game largely stops and ends at reporting. However, the effects of the toxic encounter can extend long after. Research has shown that dealing with toxicity can affect a player's long- and short-term mood and self-esteem \cite{frommelToxicityOnlineGames2024}. Yet, beyond these negative factors, when players submit a report, they hold onto expectations --- a desire for accountability and consequences. We address the question of \emph{what outcomes and expectations do players expect to get out of their reports}, especially in an \emph{ephemeral} context like games, where it can be rare to re-encounter the same toxic individual.

We view expectations and outcomes as motivations for action through an \textbf{expectancy-value theory (EVT)} lens \cite{wigfieldExpectancyValueTheory2000, wigfieldExpectancyvalueTheoryAchievement1994}, which has not yet been studied in the context of reporting systems in games. EVT frames actions as being motivated through a balance of value (the desire for specific outcomes) and expectancy (the likelihood or belief in meeting such outcomes); little has been done on whether reporting systems serve their intended purposes from the perception of \emph{what players expect}. We ground this gap in our present study around the following research questions:

\begin{itemize}
    \item \textbf{RQ1}: What values do players hold that are associated with reporting in video games?
    \item \textbf{RQ2}: To what extent does player perception of the consequences of reporting align with their desired outcomes?
    \item \textbf{RQ3}: What factors shape players' expectancy of their desired outcomes regarding reporting?
    \item \textbf{RQ4}: How do the prior factors play a role in broader perceptions of trust and expectations regarding the game?
\end{itemize}

RQ1 and RQ2 focus on values, as well as consider how players reconcile potential differences between their expectations and the outcomes that occur; RQ3 and RQ4 focus on expectancy, considering how likely users are to believe a reporting system is to provide value and how these factors influence broader perceptions regarding the game itself, the developers, and the community. 

To address these questions, we employed a mixed-methods approach. To first explore the space of reporting motivations, we created and widely distributed a formative survey to gain broad insight into player experiences and expectations. We then conducted follow-up interviews to gain a deeper understanding of players' introspective feelings and motivations for reporting. The conceptual framework of EVT shaped our analysis of the data, as we discovered that participants shared values related to both retributive and altruistic reporting motivations. Yet, players often doubted that reporting systems can fundamentally meet their expected outcomes; notwithstanding that these outcomes primarily rely on a basis of trust, which is not always present. By examining the motivations behind reporting, the primary way of addressing toxicity, we broadly understand how people expect toxic behaviour should be identified and punished, and what the desirable outcomes are --- for themselves, for the community, and for the reported perpetrator. 

\section{Related Works}

We first examine ludological research on flagging and toxic encounters to understand reporting behaviours in gaming contexts. We then expand this perspective by considering reporting within broader governance, both in real life and in online contexts. Finally, we explore the theoretical framework of EVT and its relevance to interpreting player motivations and expectations. 

\subsection{Toxicity and Reporting in Game Contexts}

Reporting in games is the primary way for players to combat toxicity, the latter being an enduring obstacle for both developers and researchers alike. The term toxicity can be broadly understood as the various negative behaviours that involve communicative abuse or disruptive gameplay behaviour \cite{beresDontYouKnow2021, kouToxicBehaviorsTeamBased2020, kordyakaCuringToxicityDeveloping2021}, and has often been studied in team-based multiplayer competitive games such as \emph{League of Legends} or \emph{Overwatch} \cite{kouFlagFlaggabilityAutomated2021a, kouToxicBehaviorsTeamBased2020, holmes2024toxicityinLeague, reidFeelingGoodControl2022}. Toxic behaviours can manifest in many ways; Kou et al. separated them into buckets of communicating aggression, cheating, hostage holding (keeping others in a negative situation), mediocriticizing (not trying hard to win), and sabotaging (consciously lowering a team's chances to win) \cite{kouToxicBehaviorsTeamBased2020}. In multiplayer games, the competitive nature and the desire for personal ranking can create frustration towards teammates; the feeling of powerlessness or control over the gameplay outcome further induces toxicity \cite{kouToxicBehaviorsTeamBased2020, zhangToxicityGameDesign2024}. Factors such as game genre, player interactions, and communication channels through text or voice can directly instigate toxicity within individual matches \cite{holmes2024toxicityinLeague, nexoPlayersDontThey2023}. Personal factors such as gender identity can also affect the perception and manifestation of toxicity in games \cite{yilmazUnderstandingGenderPC2022}. 

Although game developers often have community standards to highlight toxic actions by stating what actions are permissible or impermissible, \emph{player perception} of toxicity is subjective and contextual. Laumann \cite{laumann2021youCannotMute} and Laato et al. \cite{laatoTraumatizingJustAnnoying2024} showed that different actions are perceived differently in terms of severity, based on the underlying context and the frequency of the negative behaviour. Perception of toxicity is also affected by the broader attitudes within the gaming subculture. Beres et al. \cite{beresDontYouKnow2021} and Türkay et al. \cite{turkaySeeNoEvil2020} suggest that a level of negative behaviour has become normalized in gaming, with people viewing it as simply part of the competitive environment, to such an extent that players may even view positive communication as sarcastic rather than authentic \cite{poellerSuspectingSarcasmHow2023}. Regardless of the accepted level of toxicity, being the victim of a perceived toxic encounter within games can feel frustrating.

From the opposing perspective, researchers have investigated the reasons why players might grief and exhibit toxicity within games. These motivations often revolve around personal pleasure, for instance, by exhibiting a form of power and control over others \cite{achterboschTaxonomyGrieferType2017, juvrudROFLFckYou2020}. From a self-determination theory perspective, Achterbosch et al. showed that griefing can increase the feelings of autonomy and competence of the perpetrators \cite{achterboschAssessingImpactGriefing2024}. Researchers have considered ways to handle toxicity \cite{wijkstraHelpMyGame2023, wijkstraHowTameToxic2024}. For instance, Reid et al. developed in-game tools to support the victims of toxicity that increased their feelings of control and offered support \cite{reidFeelingGoodControl2022}. However, players often leaned towards familiar in-game tools such as muting and reporting, as they offered actionable solutions \cite{reidFeelingGoodControl2022}. 

Reporting systems are present in many online games (and digital platforms) \cite{kouFlagFlaggabilityAutomated2021a, pohjanenReportPleaseSurvey2018}. Ideally, users would report toxic behaviour that breaks the game's guidelines; these reports would lead to further punishments such as suspensions or bans \cite{kouPunishmentItsDiscontents2021}. However, while many players use reporting systems regularly \cite{pohjanenReportPleaseSurvey2018}, there remain issues regarding their perceptions and functional use. The standards for flagging differ from person to person, and unwarranted reporting of players who may not deserve it means that the system can be abused beyond its intended purpose \cite{pohjanenReportPleaseSurvey2018, kouFlagFlaggabilityAutomated2021a}. Players also question reporting's perceived effectiveness --- despite using a system meant to deal with toxicity, players do not always observe that toxic players are punished or that negative behaviour has reduced \cite{hilbersEvaluationEffectivenessSmites2015, lapollaTacklingToxicityIdentifying2020, kouFlagFlaggabilityAutomated2021a, pohjanenReportPleaseSurvey2018}. Certain systems have at times relied on crowdsourcing punishments based on reports through jury-based methods (e.g. \emph{League of Legends}' Tribunal system); people were passionate about participation but could struggle with the system or act beyond their roles \cite{kouManagingDisruptiveBehavior2017}. With more and more players, automated systems for toxicity detection and report handling have become popular in practice as well as in research \cite{blackburnSTFUNOOBPredicting2014, martensToxicityDetectionMultiplayer2015, canossaHonorToxicityDetecting2021, yangGameHateStudy2024}. 

Contrasted against prior research, our present work investigates reporting (and its tie to toxicity) at a higher meta-level. Extending on the studied in-game experiences and immediate motivations for reporting, we link these experiences towards broader motivations, the impact of reporting outside of the singular encounter, and the expected outcomes of reporting compared and contrasted against player expectations. Kou et al. highlighted how reporting behaviour represents a user-participatory subset of a broader moderation process \cite{kouFlagFlaggabilityAutomated2021a}. This wider lens allows us to understand reporting as a singular component of a more complex socio-technological process tied to community norms, player values, and system design.

\subsection{Moderation, Punishment, and Approaches to Justice}

People comply with the law for various reasons, including fear of punishment \cite{tyler2001obeying} and personal morality \cite{petrazycki2017law}. Research has shown that individuals are more likely to obey the law when they perceive it to align with their personal beliefs \cite{jackson2012people}. However, when a law is perceived as lacking legitimacy or moral justification, the mere threat of punishment is insufficient to ensure public compliance \cite{darley2009morality}. Low legitimacy can undermine people's trust and confidence in institutions such as the police or courts, leading to increased law-breaking behaviour and reluctance to cooperate with law enforcement \cite{jackson2012people}. Similarly, a lack of transparency in the legal process --- such as unfair treatment by legal authorities --- can also discourage cooperation \cite{bibas2006transparency}. In contrast, when people feel they are treated equitably under the law, they are more likely to comply with it \cite{cropanzano2015we}. One way of being complicit with the law is by reporting infractions --- in addition to self-regulation, people also report illegal behaviour. Reporting is mediated by the misconduct's severity and encouraged by social integrity \cite{khan2022examining} --- people are more likely to report misconduct when they perceive it as highly severe and in a positive ethical climate. Such environments encourage transparency and accountability, supporting individuals in taking ethical actions \cite{khan2022examining}.

To draw comparisons with other types of governance, we also examine moderation practices on non-game digital platforms, such as social media and online forums. Content moderation on these platforms plays a crucial role in addressing issues like toxicity, which can significantly impact user engagement \cite{beknazar2022toxic}. Content moderation refers to the process implemented by platform managers to oversee and filter user-generated content to protect users from harmful materials, such as misinformation, disinformation, hate speech, and online extremism \cite{roberts2017content}. This process involves determining which posts should be removed and which users should face restrictions \cite{roberts2017content}. In discussions surrounding content moderation, a critical tension arises between upholding freedom of expression and preventing harm caused by misinformation \cite{kozyreva2023resolving}. This moral dilemma has been explored in prior research, which found that most people prioritize harm prevention over protecting free speech \cite{kozyreva2023resolving}. This preference may stem from concerns about the negative impacts of misinformation on societal well-being.

While automated tools play an essential role in content moderation, user reporting remains a critical component \cite{mcinnis2021reporting}. Users report others for various reasons, including perceived violations of platform rules or personal grievances \cite{evans2012governing}. While user reporting serves as a vital mechanism for flagging potential issues, its effectiveness and fairness are closely tied to broader principles of transparency and accountability, such as those outlined in the Santa Clara Principles \cite{santa_clara_principles}. In 2018, several organizations, advocates, and academics launched and endorsed the Santa Clara Principles, which are a set of guidelines focused on transparency and accountability in online content moderation to support and empower users \cite{rosal2020contribution, santa_clara_principles}. The guideline included several core elements, including notices, explanations and appeal mechanisms --- platforms must inform users when their content is removed or restricted, clearly state which rule was violated and how it applies to the removed content, and provide users with a fair and transparent way to appeal moderation decisions. We extend upon these principles while focusing specifically on user reporting as a mechanism for moderation and its implications for governance on multiplayer gaming platforms.

\subsection{Theory of Motivations - Expectancy-Value Theory} 
There are many different frameworks of motivations that have been explored in academic ludology. For example, self-determination theory, which buckets psychological needs into those satisfying autonomy, competency, and relatedness, has often been used as the lens to investigate why people are attracted to gaming \cite{deciSelfDeterminationTheoryWork2017}. Ryan et al. show that such feelings of autonomy and competence during gameplay are associated with personal enjoyment and preference \cite{ryanMotivationalPullVideo2006}. However, whereas self-determination theory explains psychological motivations behind a sustained activity over time, we considered expectancy-value theory (EVT) as a suitable framework for investigating instantaneous, one-off decision-making processes such as reporting. 

Under EVT, motivation to perform a task is mediated by the concepts of \emph{expectancy} and \emph{value} \cite{wigfieldExpectancyValueTheory2000, wigfieldExpectancyvalueTheoryAchievement1994, wigfieldWhatDoesExpectancyvalue2019}. Value can be broadly construed as the desire for specific end states, and Wigfield and Eccles' work has categorized values into those of \emph{attainment value} --- the importance of doing well on the task, \emph{intrinsic value} --- the enjoyment one gets from the task, and \emph{utility value} --- the importance of the task for the future \cite{wigfieldExpectancyValueTheory2000, wigfieldExpectancyvalueTheoryAchievement1994}. Tasks also instill a cost that detracts from value, which comprises the time and effort of the task, the external effort to advance the task, opportunity cost for alternatives, and changes in emotional affect \cite{ecclesExpectancyvalueTheorySituated2020, flakeMeasuringCostForgotten2015}. Expectancy refers to the expectations, or \emph{belief}, that behaviour can lead to the desired outcomes \cite{wigfieldExpectancyValueTheory2000, ecclesExpectancyvalueTheorySituated2020}. EVT has largely been employed in educational and developmental contexts to understand the motivations for achievement \cite{wigfieldWhatDoesExpectancyvalue2019, nagengastWhoTookOut2011a, putwainExpectancySuccessAttainment2019a}. However, it has been studied more broadly as well, such as in game contexts. Rachmatullah et al. found that outcome-expectancy belief was a positive predictor of flow experiences in games, whereas lower self-efficacy (personal belief) resulted in increased feelings of frustration \cite{rachmatullahModelingSecondaryStudents2021}; Lin et al. highlighted the importance of achievement value on player engagement in massively multiplayer online role-playing game (MMORPGs) \cite{linUnderstandingPlayersAchievement2015}.  

EVT provided a suitable framework for our work because our research questions strongly focus on expectations and outcomes --- we explore factors such as what values people place on the outcomes when deciding to report in games, mediated by the likelihood of such consequences. Our work explores how the mental calculations core to EVT are affected by discrepancies between expectations and observed experiences, transparency of the process, and the alignment of norms. 

\section{Methods}

To investigate our research questions, we performed a mixed-methods approach to data collection and analysis. We generally followed an explanatory sequential design \cite{dawadiMixedMethodsResearchDiscussion2021} --- starting with an early-stage, formative exploration of quantitative trends that informed deeper qualitative analysis to understand these trends. We started broadly, deploying a survey to garner broad initial sentiments regarding motivations and expectations behind reporting practices. Motivated by the survey results, we dug deeper into specific experiences and feelings through rich interviews with video game players. Our methods were reviewed and approved by our institute's ethical review board. 

\subsection{Survey}

\subsubsection{Survey Development and Procedure}

The goal of the survey was to broadly explore the range of reporting habits in games, bringing together a wide range of participants from a wide range of games. We used this data to understand patterns in motivations for reporting, as well as people's general expectations for its outcome. The initial design of the survey was motivated by prior research and researchers' brainstorming --- taking the view of researchers as experts, as all researchers had extensive experience with playing video games and using their reporting systems. Although our survey touched on questions later interpretable through EVT, such as the value of reporting (e.g. community-improvement, avoiding toxic players) and its potential expectancy of reporting (e.g. chances of consequence), the theoretical lens of EVT was not imposed during survey construction. Rather, we wanted to broadly understand the space and the most general trends around reporting behaviours. 

To highlight these trends and to provide a level of explainability around the behaviours, the survey consisted of a mix of multiple-choice, Likert-scale and open-ended questions. The required Likert-scale questions asked participants to rate their agreement with statements regarding reporting expectations, motivations, and outcomes; the optional open-ended questions mainly offered space for participants to freely expand upon their ratings with their rationale. 

To begin, we asked participants to select behaviours they would report for and what punishment they expect for each behaviour, with the toxic behaviours inspired by Kou's prior classification \cite{kouToxicBehaviorsTeamBased2020}, and the punishment outcomes based on our own gameplay experiences combined with Ma's classification of punishment types \cite{maTransparencyFairnessCoping2023a}. To understand experiences around reporting, we asked participants to rate their agreement with motivations for reporting around various mediating factors, e.g. understanding of transparency, mental model of consequences, and potential for reformation. These were inspired by the researcher's readings of prior work in both game moderation \cite{maTransparencyFairnessCoping2023a, kouPunishmentItsDiscontents2021,kouFlagFlaggabilityAutomated2021a} and broader moderation in social media \cite{zhangCleaningStreetsUnderstanding2024a, maHowUsersExperience2023}. For instance, Zhang et al. \cite{zhangCleaningStreetsUnderstanding2024a} highlighted how participants may have differing understandings and motivations for reporting on social media, and we translated these questions into a games context. Zhang et al. \cite{zhangCleaningStreetsUnderstanding2024a} also raised how reporting is both a collective duty and an individual right, which became questions regarding the extent to which reporting in games is driven by collectivist outcomes or personal desires. Ma et al. \cite{maTransparencyFairnessCoping2023a} highlighted how transparency and fairness impact moderation in gaming; we asked questions regarding the extent to which punishment transparency might also play a role for the reporter.


The full survey is in the supplementary materials. Internal face validation was conducted through pilot testing --- we asked HCI researchers with game expertise to subjectively critique our survey structure, identify confusing questions, and assess coverage of our RQs. We estimated that the survey would take 30 minutes to complete, and we offered compensation of \$6 CAD. 

\subsubsection{Survey Participants}

We distributed the survey online through Prolific\footnote{\url{https://www.prolific.com/}}, an online crowdsourcing platform for user studies. In the survey posting, we highlighted that the main inclusion criteria were to have experience with video game reporting systems (which we also added as a screening question in the survey itself). We also pre-screened participants based on their approval rating and the number of previous submissions, aiming for higher-quality data. However, after data collection, we further cleaned the responses to remove responses with implausible fast completion times (under 4 minutes), incomplete surveys, and non-English responses, arriving at a final sample of 98 participants. The ages ranged from 18 to 62 (average: 28.5), and the gender distribution was 62 male, 35 female, and 1 non-binary. In terms of gaming experience, 62 participants had played games for over 10 years, 25 had played games for 5--10 years, 6 had played games for 2--5 years, and 5 had only played games for less than 2 years. In terms of frequency, 30 participants played games for more than 10 hours a week, 34 participants played 5--10 hours, 21 played 2--5 hours, 11 played less than 2 hours, and 2 participants indicated that they currently are no longer playing games. We found the demographics of our survey participants to be acceptable, providing a satisfactory sample of gaming experiences and frequencies from different age ranges. 

\subsection{Interview}

\subsubsection{Interview Protocol Procedure}

The interviews chronologically followed the surveys, after we had already started initial data analysis and highlighted initial patterns. The goal of the interview was to delve deeply into the experiences and rationale underlying the trends in motivation and expectations from the survey. The interview provided a deeper explanation and reasoning for these trends, as well as exceptions and counterexamples. The interviews asked about people's reporting experiences, what about the encounter stuck with them, and the relationship between reporting and community improvement. 

The qualitative, flexible nature of the semi-structured interview allowed us to understand experiences at a broader scope compared to the questionnaire --- for instance, instead of ranking agreement with existing reasons for reporting, we could ask participants to provide their existing rationale unbounded by our prior expectations. We view the interview as still exploratory --- exploring the space of reporting habits, their experiences, and their expected outcomes. In this spirit, our interview questions were related to EVT constructs but still open-ended, aiming to understand their rationale for reporting and what kind of outcomes they expected and why (e.g. ``Why do you report people?'', ``What is reporting meant to do?'', etc.), even though follow-up questions could diverge into broader exploration. We conducted initial pilot testing with HCI researchers to evaluate the subjective face validity of our interview questions. 

The interview protocol can be found in the supplementary material. The interviews took approximately 40--60 minutes online over Zoom, and each participant was reimbursed \$16 CAD. Audio recordings were collected with participant consent. 

\subsubsection{Interview Participants}

Participants were recruited from a listing on our institute's paid studies page, with the main inclusion criteria being having experience with reporting others in games (in addition to being fluent in English). We recruited a sample of 19 participants spanning a wide variety of different genres, gaming experiences, and reporting habits --- some participants had only recently started playing games and had only reported toxicity a couple of times, others had been playing multiplayer games for over a decade and issued a multitude of reports. The researchers subjectively felt that this sample of participants provided sufficient information power \cite{malterudSampleSizeQualitative2016c}, based on factors including the specificity of experience, the quality of interviews, and so forth; this sample size also exceeds local standards for HCI \cite{caineLocalStandardsSample2016b}. Information power was subjectively judged by the primary researcher on an ongoing basis during the interviews (in discussions with the research team); we stopped collection when we found that additional data were yielding diminishing new insights into answering the research questions. 

One factor that might cause people's experiences to differ is the genre of games that people play and their experience with games as a whole. As such, before the interview, participants were asked to send the researcher information about the games they had played and reported people in (the researcher read up on the game in the cases where they were not familiar), their frequencies of reporting, and whether they had been punished based on other reports. We kept the latter two questions unstructured so participants could self-report their diverse and unique experiences, and participant responses to them can be found in the work's supplemental material. 

Furthermore, during the study, we collected information regarding participants' age, gender, frequency of video game play, and their years of experience playing games (which we summarized in Table \ref{table:demographics}). The average age of the participants was 24.3, the gender distribution was 9 male, 9 female, 1 non-binary, and most of the participants were experienced gamers with over 10 years of experience. Although a wide range of games were mentioned, the majority of games discussed in-depth were team-based multiplayer games (such as \emph{League of Legends}, \emph{Overwatch}, \emph{Valorant}).  

\begin{table*}
\centering
\caption{Summary of Interview Participants}
\renewcommand{\arraystretch}{1.5}
{\rowcolors{2}{white}{mypink2} 
\begin{tabular}{c c c C{6cm} C{2cm} C{2cm}}

\hline
\textbf{ID} & \textbf{Age} & \textbf{Gender} & \textbf{Games Discussed} & \textbf{Frequency of Play (Weekly)} & \textbf{Years of Experience}  \\ 
\hline
    \noalign{\vskip 1mm}   
    P1 & 21 & M & World of Tanks Blitz, World of Warships, Clash of Clans, Battlefield 2 & <2 hours & >10 years \\ 
    P2 & 27 & M & League of Legends & 5-10 hours & >10 years \\ 
    P3 & 29 & M & League of Legends, Overwatch & 2-5 hours & 2-5 years \\ 
    P4 & 25 & M & League of Legends, Overwatch, Fortnite & 2-5 hours & 5-10 years \\ 
    P5 & 30 & F & Teamfight Tactics, Clash Royale & >10 hours & >10 years \\ 
    P6 & 23 & F & League of Legends, Valorant, Club Penguin & 5-10 hours & 2-5 years \\ 
    P7 & 22 & M & Valorant, Rainbow 6 Siege, GTA Online, Chess.com & 2-5 hours & 5-10 years \\ 
    P8 & 20 & F & Golf Battle & 5-10 hours & >10 years \\ 
    P9 & 21 & NB & Minecraft, Roblox, Identity V & 2-5 hours & >10 years \\ 
    P10 & 19 & M & Valorant, Overwatch, Fortnite  & 5-10 hours & >10 years \\ 
    P11 & 18 & M & Rainbow 6 Siege, PUBG, Rocket League, Counter-Strike 2 & >10 hours & 2-5 years \\ 
    P12 & 23 & M & Call of Duty Mobile & 5-10 hours & >10 years \\ 
    P13 & 20 & F & Fortnite, Splatoon & 5-10 hours & >10 years \\ 
    P14 & 38 & M & Dota 2 & >10 hours & >10 years \\ 
    P15 & 18 & F & Valorant, Roblox, PUBG & <2 hours & >10 years \\ 
    P16 & 27 & F & Deadlock & 5-10 hours & <2 years \\ 
    P17 & 32 & F & Overwatch & 5-10 hours & >10 years \\ 
    P18 & 23 & F & Valorant, League of Legends, Counter-Strike 2 & >10 hours & 5-10 years \\ 
    P19 & 25 & F & Valorant & 5-10 hours & 5-10 years \\ 

    \noalign{\vskip 1mm}   
\hline

\end{tabular}}
\Description{Table illustrating the demographic data of the interview participants, including age, gender, games discussed, and frequency of gameplay}
\label{table:demographics}
\end{table*}

\subsection{Data Analysis}

The data analysis process focused on the concept of triangulation \cite{carterUseTriangulationQualitative2014a} --- using multiple methods to understand the reporting phenomena. We first aimed to understand reporting phenomena broadly, before narrowing into deeper interpretation. We started with our collected survey data, which consisted of both quantitative and qualitative data. Our quantitative data was largely exploratory --- we designed the questions around understanding trends in the motivations and expectations around reporting patterns. We inspected the quantitative data through exploratory data analysis to understand these trends at a high level. However, given the survey's formative nature, we did not conduct any specific statistical tests on the quantitative data. 

Still, we used patterns from our quantitative exploration to inform our qualitative analysis of the open-ended survey responses, employing a thematic analysis approach \cite{braun2021TA}. The first and second authors collaboratively began with initial coding on the data, generating code labels that were iteratively refined as the process continued. Visual mapping approaches were used to develop relationships and categorize the codes, which later informed the development of themes. Although the coding was done independently, both authors helped generate and discuss code labels, relationships, and themes. These themes formed initial inductive trends regarding reporting practices. Even though our collected data from the survey was broadly related to constructs of motivations, value, and expectancy, it was during data analysis that the theoretical lens of EVT sharpened. When analyzing the data, we found that EVT provided a rational framework for us to view and articulate why people were motivated to report (i.e. the value from reporting), and what kind of consequences people wanted from reporting (i.e. the expectancy of reporting). 

Consequently, the EVT framework helped shape both the structure and analysis of our interviews. Although the questions in our interview were influenced by our findings from our survey, the analysis of the interviews began with an initially inductive approach, understanding the data as a whole without prior assumptions. This was done to check whether there were findings that emerged beyond the initial questionnaire findings or the broader theoretical EVT framework. The first and second authors again started with initial exploratory coding to understand the experiences and perspectives of the participants. In this initial exploratory coding, and ongoing throughout the analysis process, we engaged in ongoing discussion about the data, the data labels, and their own thoughts and interpretations of the data (which was also discussed regularly with the team).  

As the analysis progressed, we refined and organized the initial codes. This process of refinement was iterative, driven through comparisons across participant data and discussion among the researchers when codes were too granular or too broad to be effectively applied. After this coding process, we shifted towards a more deductive, theory-driven perspective to translate the relationships in our code labels into defined themes. EVT informed the mapping and grouping of the codes, as we developed relationships and hierarchies in the data (see supplementary material); this process later crystallized into themes around the ``expectancy'' and ``value'' of reporting. For instance, the codes ``Reporting for Spite / Revenge'' and ``Reporting for Altruistic Reasons'' highlight the split motivations that interview participants spotlighted, which became grouped under the EVT construct of ``People's Values''. On the other hand, codes like ``Feeling like the Developer Cares'' and ``Communication of Reporting Feedback'' tied to how participants shaped consequences for their reports, and became grouped under the EVT construct of ``People's Expectancy''. 

As our interview data extends our survey data through richness and depth, we constantly made links between the two sources as we progressed deeper into the analysis. We considered how patterns from the interview data could both fit into and challenge existing trends we had highlighted previously. Given the rather formative and early-stage approach of our questionnaire (especially as not everyone opted to answer every open-ended question), we mainly lean on qualitative quotes from our rich interviews in the findings. The survey data follows similar trends, and we use the quantitative survey data as support in presenting our findings.

\section{Findings}

\subsection{\emph{RQ1: What values do players hold that are associated with reporting?}}

\begin{figure*}[h]
  \centering
  \includegraphics[width=0.9\linewidth]{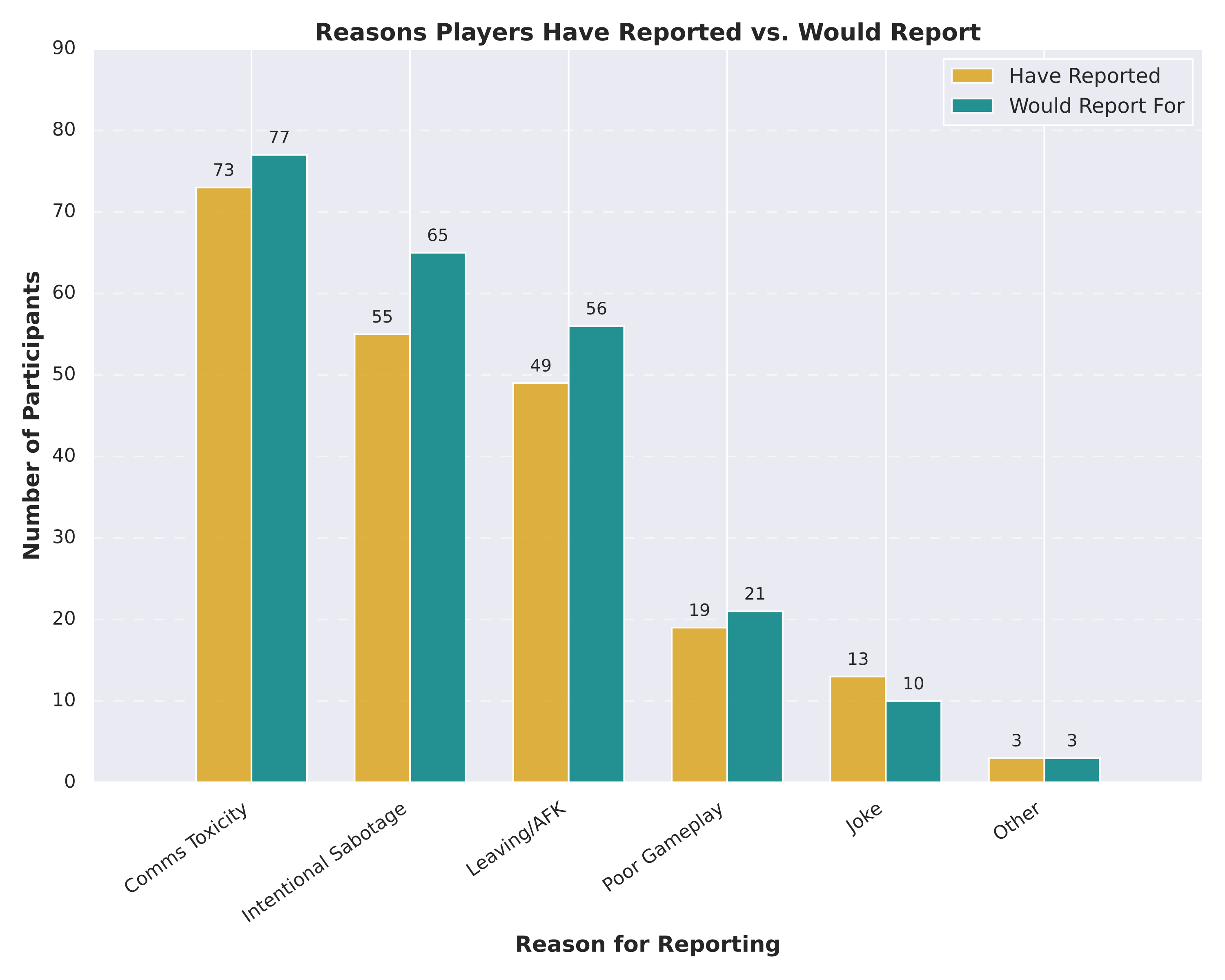}
  \caption{Number of questionnaire respondents that have reported and would report for specific reasons (out of 98 respondents total)}
    \Description{Grouped bar chart depicting people's experienced and hypothetical reasons for reporting, including reasons such as comms toxicity, intentional sabotage, and so forth.}
  \label{fig:reportingreasons}
\end{figure*}

We outline the values regarding reporting --- subjective task value is constructed by potential benefits weighed against costs \cite{muenksCostSeparatePart2023, ecclesExpectancyvalueTheorySituated2020}. People report for different gameplay reasons (illustrated in Figure \ref{fig:reportingreasons}, with grouping inspired by Kou \cite{kouToxicBehaviorsTeamBased2020}). However, how people's perceived task \emph{values} for reporting are different depending on their motivation, which we initially explored through the questionnaire responses (see Table \ref{table:reasons}) but expanded on during our interviews. We identified two main categories of motivations driving distinct benefits: \textbf{retributive} motivations --- of wanting to `get back' at the reported player and \textbf{altruistic} motivations --- of wanting to improve the community for everyone. Some of the interview participants were more readily influenced by one set of such motivators in reporting, but the vast majority indicated a mixture of both.

\begin{table*}
\centering
\caption{Agreement levels to the questions regarding motivations for reporting from 1 (not a motivation at all) to 5 (a strong motivation) from our questionnaire [n=98]}
\renewcommand{\arraystretch}{1.5}
{\rowcolors{2}{white}{mypink2} 
\begin{tabular}{c C{7.5cm} c c}

\hline
\textbf{Question ID} & \textbf{Reason} & \textbf{Mean} & \textbf{Std. Dev} \\ 
\hline
    \noalign{\vskip 1mm}   
    Q1 & To make the game a better place for everyone & 4.59 & 0.82 \\
    Q2 & To prevent future toxic behaviour and improve the quality of gameplay for others & 4.62 & 0.68 \\
    Q3 & To feel a sense of self-satisfaction or closure after experiencing negative behaviour & 3.10 & 1.40 \\
    Q4 & To avoid encountering negative behaviour in future games for myself & 4.43 & 0.85 \\
    Q5 & Because it is part of my responsibility as a player to uphold the community standards & 3.69 & 1.20 \\
    Q6 & To hold players accountable for their actions & 4.36 & 0.94 \\
    \noalign{\vskip 1mm}   
\hline

\end{tabular}}
\Description{Table of questions related to the motivations of reporting, along with their mean and standard deviations along a 5-point Likert Scale}
\label{table:reasons}
\end{table*}

\subsubsection{Benefits Associated with \textbf{Retributive} Motivations}

Participants discussed how part of the motivation that comes with reporting is retributive --- wanting to take their anger out and blame someone after having a negative experience. We interpreted this motivation to be highly influenced by the player's own temporary experience within the game, such as the match outcome or the player's temporary mood. For instance, P2 mentioned that sometimes after losing a match in \emph{League of Legends}, players might feel the need to:

\begin{quote}
    \emph{``Let out frustration ... you're looking for someone to blame almost, I would say, especially right after you've lost''} - P2
\end{quote}

Retributive motivations tied rather closely with immediate intrinsic value --- players used reporting as a way to feel better in the short-term, as doing so could provide \emph{``a sense of catharsis in hitting the button''} (P14) and \emph{``revenge ... you've been toxic to me so I'm going to report you so you can't speak next game''} (P19). The reporting system provides the primary way for a player to vent out their negative emotions, as well as a way of regaining control. In multiplayer team games, players felt a lack of control as they were forced to engage with this negative player for the full duration of the game --- \emph{``I feel like toxicity in that context you can't really control ... because League of Legends each game is 30-35 minutes''} (P2). Through reporting, players regained this control --- P7 states that the reporting player exerts \emph{``some amount of power''} because of \emph{``the ability to get someone penalized''}.

Penalties tie into the utility value behind retributive motivations --- the outcome that people desire from reporting from this perspective is the punishment of the reported player. In the ephemeral context of games, where it can be rare to run across the same players in most ranks of gameplay, we interpreted this actual outcome as rarely offering any actual long-term self-benefit --- whether someone you never see again is punished or not does not truly matter. P2 extended on their prior statement, revealing that \emph{``deep down, you know that it's not really a rational thing that you're doing''}. 

Since retributive motivations are impacted by mood and outcome, reports made with such motivations may not always have the best intentions. Most participants spoke out against reporting people for invalid reasons, such as \emph{``if someone has a better KDA [author note: KDA = Kills/Deaths/Assists, a common metric for performance in multiplayer games] than them''} (P15). At the same time, several participants mentioned that they sometimes do use the report button for reasons that admittedly may not be deserved:

\begin{quote}
    \emph{``Sometimes I used to report people for fun, if I lost or something''} - P12
\end{quote}

\begin{quote}
    \emph{``If I get tilted and then I see someone's not doing too well even though they might just be not having a good day''} - P10
\end{quote}

Thus, even though reporting is primarily viewed as a way to combat toxicity, it can be used in a toxic manner as well --- potentially ruining someone else's experience due to personal frustrations. 


\subsubsection{Benefits Associated with \textbf{Altruistic} Motivations}

The other main motivator associated with reporting was altruism --- wanting to improve the gameplay experience and the broader community for everyone. This motivation was influenced by the player's personal values, their perceptions of accountability, and their moral judgment. P9 stated the role of reporting was for \emph{``making it better for the next player''} and P6 ties it towards \emph{``fair[ness] and justice''}. Reporting behaviour in this case was mediated by the perceived intention of the other player towards not just themselves, but towards others as well. From this perspective, the reporting system serves as the primary way to combat negative behaviour and make a difference in the community. P18 highlighted their experience with sexist comments in video games, reporting to help others in the future:

\begin{quote}
    \emph{``There's not a lot of girls that play ... so when I report someone, if a girl starts playing ... she doesn't have to experience that''} - P18
\end{quote}


We associate this idea of wanting a better community with the utility value of altruistic reporting. Participants were aware that one report does not change the community. Thus, in some cases, there is no immediate altruistic intrinsic value for the players themselves --- as P13 mentioned \emph{``talk and things like that, they don't personally bother me that much, I'll normally just mute and move on''}, highlighting how, for their immediate experience, dealing with the minor annoyance is simple even without reporting. However, P13 continued --- \emph{``I feel like it's my responsibility to report actions ... that there's not some younger girl on the other side of the screen that will experience that and have that affect them''}. 

For most participants, the core altruistic value for reporting is in cultivating a positive environment for the future, beyond the ephemeral, toxic encounter. This is not solely for the benefit of others, as the reporting player also wants to experience an improved, less-toxic environment. This tied into the personal value that certain players saw in games as a whole:

\begin{quote}
    \emph{``Gaming is to build a community of people you would probably game with all the time ... being rude towards somebody ... takes away from the meaning of it being a fun thing''} - P16
\end{quote}


\subsubsection{Benefits are Mediated by the Dimensions of Specific Games}

The benefits of reporting are unique to specific game contexts; several factors affected how players perceived reporting value within specific games. Two factors discussed previously were the ephemerality of interaction --- where it is rare to see the same player in different matches in multiplayer team-based games, and the forced temporal interaction --- where players are forced to share the gaming space with others for a set amount of time. However, participants also brought up how factors such as team dynamics and cooperation requirements that could induce toxic reportable behaviour, e.g. \emph{``when someone isn't playing with the team then that makes me want to report them''} (P10); this was exacerbated by competitive settings \emph{``in competitive games, the frustration is usually higher ... so reporting and other stuff is more [often]''} (P11). Communication channels within games were also discussed, especially regarding the impact of voice chat: \emph{``it makes people more toxic because it's easier to talk shit to your teammates''} (P10). When considering the list of discussed games in Table \ref{table:demographics}, there is some potential bias towards multiplayer team-based games. However, people's responses suggest that reporting is more prevalent in these types of games precisely because of these contextual reasons. 



\subsubsection{Costs Associated with Reporting}
Under EVT, task value increases through the benefits of an action but decreases as a result of its costs \cite{ecclesExpectancyvalueTheorySituated2020}. In the context of reporting, however, the \emph{cost} associated with reporting was generally perceived to be low. Participants indicated how games typically make it extremely easy and low effort to put in a baseline report, making it accessible to any person --- \emph{``you just have to click a couple buttons''} (P10). This relatively low cost of reporting also aligns with our survey response for the question \emph{''it requires a great deal of effort for me to report a player``} (mean=2.84, std=1.39).

A reporting system that would incur more effort and a higher cost might be less attractive for people to use, e.g. \emph{``if I had to manually go to a website or form I would probably not report almost anyone''} (P17). However, a low cost can also be a problem given our prior discussion regarding reporting as a potentially abusable behaviour --- P2 mentioned that it is precisely because of this low cost that they see it as \emph{``kind of a joke just because you can just do it with a few buddies with no real evidence needed''}. 

Even though the baseline cost for submitting a report is low, games often provide the opportunity to add text as supporting evidence. Some participants were willing to incur this added effort to obtain the aforementioned value, e.g. P18 mentioned that with a \emph{``free text option I'd be writing to the max characters''}. Tying into the retributive rationale for reporting, P2 mentioned that the ease of reporting means that they can easily attain intrinsic satisfaction during the game, \emph{``just take a pause to report the person and it's like a minor cathartic moment''} (P19).

\subsection{\emph{RQ2: To what extent does player perception of actual consequences of reporting align with expected outcomes?}}

Building upon participants' expectations of reporting outcomes, which differed based on their underlying motivation, we examine the extent to which players believed reporting could lead to these desirable outcomes. Specifically, we consider whether players perceive reporting as an effective means to (1) punish offenders and (2) foster a better community. 

\subsubsection{Reporting and the Expectation for Player Punishment}

The retributive outcome associated with reporting is for the reported player to be punished. However, all players were cognizant that not every one of their reports led to the desired effective punishment against the player. Other than the most egregious or easily detected cases, participants indicated how game moderation systems typically punish reported players after a sustained pattern of wrongdoing. Participants had mixed attitudes towards this, with some being accepting of this leniency --- \emph{``maybe they should be given a chance''} (P5). However, others felt more strongly about punishment after a single offence:

\begin{quote}
    \emph{``If you grief one [game], you're still harming a lot of people's experiences''} - P10
\end{quote}

\begin{quote}
    \emph{``Letting people get away with it once because they've never done it ... kinda defeats the purpose of the system''} - P2
\end{quote}

For these cases, people were focused on their report as correctly identifying toxicity, so in that case, reporting \emph{should} logically lead to punishment. Yet, despite their internal beliefs, we have already established that this cannot hold for every single report. We have acknowledged that the report button can be used for the wrong reasons, highlighting the importance of added contextual information --- \emph{``one report against a player who was never reported before could be for something severe, or it could not be''} (P16). Although participants \emph{speculated} that not all reports were punished, the lack of transparency in typical report systems makes it hard for them to \emph{confirm} --- we extend upon reasons on why participants held such beliefs when we discuss outcome \emph{expectancy} in the next findings section.

Even when the reported players were punished, players were ambivalent about the structural factors of the punishment. Several participants indicated that they would like to see more severe punishments to act as a deterrent; the survey responses also provided loose support to this: \emph{``I believe that current punishments for my reports are too light''} (mean=3.26, std=1.06). This was mainly tied to the ease of circumventing the report:


\begin{quote}
    \emph{``I don't think any of it is getting taken seriously ... people can just open a new e-mail address and a new gaming profile''} - P12
\end{quote}

Returning to retributive values, players sought to punish the actual reported player (not their account) and attain revenge, but the ease of circumvention, especially in free-to-play games, makes the current consequence misaligned with the desired outcome. Players did note that harsher punishments could be employed to ensure the actual punished player is not allowed to continue playing, such as banning by IP address, banning based on hardware IDs, or account subscription fees, but those come with problems as well, e.g. \emph{``people say IP banning could ban their siblings' account by accident''} (P5). 

\subsubsection{Reporting and the Expectation for a Better Community}

The altruistic outcome associated with reporting is in fostering a better, less toxic community. As the short-term outcome of reporting is potential punishment, we begin by understanding what players consider to be the relationship between punishment and an improved community. Firstly, whether punishment was an effective way of quarantining toxicity was considered, however, we have already identified that the ease of making new accounts limits the effectiveness of punishment in this consequence, e.g. \emph{``people get banned and make new accounts and ... continue the cycle all over again''} (P18); in such a way, toxicity does not improve unless people are encouraged through punishment to change their behaviour. 

Some participants were optimistic about punishment to deter toxicity and underscore change:

\begin{quote}
    \emph{``Some people will be afraid of being temporarily banned or being punished. So at a point I think they will, they will behave themselves''} - P4
\end{quote}


People may want to keep playing the game or have a value associated with their account that they do not want to lose, e.g. P11 shares an anecdote where \emph{``one of my friends he's very toxic sometimes''}, yet after getting banned, \emph{``he was forced to turn off team chat, so it kind of works sometimes''}. Thus, punishment can improve the community when people are incentivized to reflect on their actions and change, even if it is solely so that they can keep playing. 

However, others were less optimistic about punishment as changing player attitudes --- P2 states that games are \emph{``not the kind of setting that people want to be reformed in''}, P10 extends that people's toxic habits only change if:

\begin{quote}
    \emph{``The person already wants to change, but otherwise no, I don't think it'll help make that person a better person''} - P10
\end{quote}

If people cannot reflect on their actions and change after punishment, then reporting does not help in making the community a better place, as the person simply returns and continues to act similarly. As P17 mentions \emph{``I don't think [punishment] necessarily fixes or makes [the reported player] reflect''}. Despite this, players had difficulty identifying what alternatives could reach this outcome:

\begin{quote}
    \emph{``Punishment isn't the best but you can't administer therapy for everyone ... punishment isn't a great deterrent and we know it doesn't really work. I wouldn't know what an alternative is''} - P18
\end{quote}

This is partially due to factors beyond reporting (and thus out of the player's control), e.g. P6 mentions how, even as toxic players get punished, there is always an unending battle because \emph{``there's always an influx of players''}. 


\subsection{\emph{RQ3: What factors shape players' expectancy of their desired outcomes?}}

\begin{table*}
\centering
\caption{Agreement levels to the questions regarding expectancy variables from 1 (Strongly disagree) to 5 (Strongly agree) [n=98]}
\renewcommand{\arraystretch}{1.5}
{\rowcolors{2}{white}{mypink2} 
\begin{tabular}{c C{7.5cm} c c}

\hline
\textbf{Question ID} & \textbf{Statement} & \textbf{Mean} & \textbf{Std. Dev} \\ 
\hline
    \noalign{\vskip 1mm}   
    Q7 & My reports typically lead to the consequences I expect & 3.32 & 0.88 \\
    Q8 & I am typically satisfied with the outcomes of my reports & 3.39 & 1.02 \\
    Q9 & I am motivated to keep reporting players even if the outcomes do not always match my expectations & 3.93 & 0.96 \\
    \noalign{\vskip 1mm}   
\hline

\end{tabular}}
\Description{Table of questions related to the expectancy of the reporting task, along with their mean and standard deviations along a 5-point Likert Scale}
\label{table:expectancies}
\end{table*}

\begin{figure*}[h]
  \centering
  \includegraphics[width=0.95\linewidth]{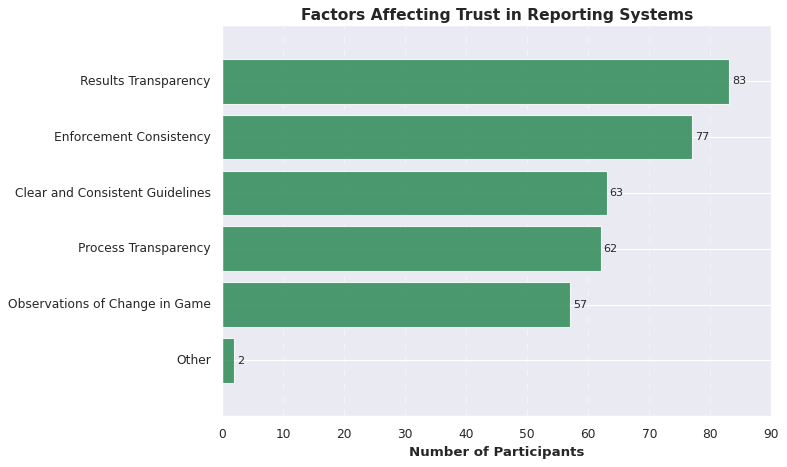}
  \caption{Number of questionnaire respondents that agreed with how much each factor affected trust (out of 98 respondents total)}
  \Description{A horizontal bar chart on the frequency of how various factors of results transparency, guidelines, and other factors affected respondent trust in reporting systems. }
  \label{fig:trust}
\end{figure*}

We highlighted how the bridge between reporting action and resultant outcomes can be disconnected from the results from both retributive and altruistic perspectives. Here, we explore the factors that influenced why (or why not) players may \textbf{expect} their desired outcomes to be achieved. From Table \ref{table:expectancies}, we find that survey respondents reported neutral-to-positive scores on direct expectancy (Q7) and outcome satisfaction (Q8). From Figure \ref{fig:trust}, factors such as transparency in process and results, clarity of reporting guidelines, consistency of perceived enforcement, and observations of improvements in the game over time all comprise significant factors that comprise feelings of a player's \emph{trust} in the system. 

We found the concept of \emph{trust} to be a major factor in affecting expectancy, appearing commonly in participant narratives and our data analysis. While trust is not the sole factor that impacts expectancy, it plays a major part in a reporting context simply due to the non-transparency of the system --- players must hold a level of trust that the black-box reporting system operates correctly; as we find, this is mediated by trust in more tangible aspects, i.e. the developer and their fellow players. Trust is especially important in uncertain, opaque conditions, and can be shaped by personal views and experiences of external facets ``around'' an opaque system, such as the institution or company using that system \cite{steedmanComplexEcologiesTrust2020}. We find that trust, along with other a priori construals, underpin the more EVT-aligned construct of individual \emph{belief} \cite{ecclesExpectancyvalueTheorySituated2020} that reporting will provide meaningful value.

\subsubsection{Belief is Affected by Experiences}

Participants' beliefs that their report would lead to a desired outcome were driven by their past and ongoing experiences. This comprised both previous knowledge of the developer and their approaches to toxicity, as well as the current perception of the effectiveness of specific reporting systems.  

\paragraph{Prior Experience with Developers}

The extent to which players perceived the developers' attitudes toward the game significantly influenced their expectancy related to reporting outcomes. If players trusted the game developers, they would hold a stronger expectancy of attaining desired outcomes. In contrast, when players believed addressing toxic in-game behaviour was a low priority for the developers, they had low expectancy of attaining their desired outcomes. Part of the trust in the developers came from the players' prior perception of reputation. P14 highlighted this perspective: 

\begin{quote}
    \emph{``[Valve] [author note: Valve is a video game developer and publisher, notable for the Dota and Counter Strike series of games] is seen as a company that does things the right way and have earned the trust of the community... it extends into all other facets of how they manage their games"} - P14
\end{quote}

One way of establishing this perceived trust and rapport is through visible community engagement:

\begin{quote}
    \emph{``If a game developer is really involved with the community... it usually brings a lot of generally good people, positive people to the game as well"} - P7
\end{quote}

Developer actions play a major role in influencing their reputation, the perception of the community, and player engagement with the entire reporting ecosystem. 

On the contrary, a low level of trust in developers can cause players to consequently fail to believe in the reporting system as having any value. For example, P8 had a negative perception of the developers of \emph{Golf Battle}, stating that they prioritized short-term profits over creating a sustainable gaming environment:

\begin{quote}
    \emph{``It’s the case with many mobile games. The purpose is more about in-app purchases and profits for the company rather than making a successful long-term game”} - P8
\end{quote}

When players negatively perceive the game developers for other reasons, they subsequently distrust the game developers to handle reports as well --- P8 expanded that \textit{``You don’t expect anything from the developer, and the developer doesn’t really expect anything from the community either”} (P8). P2 shared a similar opinion, expressing that when players feel ignored by developers, they lose confidence in the developers’ ability to address all issues effectively, stating that toxicity was:

\begin{quote}
    \textit{``An issue that they could have really fixed if they wanted to a long time ago, but went neglected for so long that at this point it's too late"} - P2
\end{quote}

P2 cited how this loss of trust caused him to also lose belief in the efficacy of the reporting process \textit{``you can't trust the developers that they're not punishing the right people or giving the right punishment for certain people}". The loss of trust undermines player expectancy in the reporting process and contributes to a less supportive gaming environment.

\paragraph{Ongoing Experience with the Game}

While dealing with toxicity, players were still engaged in an ongoing experience of playing the game, and one factor that shaped their expectancy was their observation of toxicity within the game over time. Whether the desired patterns of changes (of player punishment or community improvement) were actually observed fundamentally shaped expectancy, even if it was reliant on anecdotal evidence or was a result of the broader, more external ecosystem rather than solely reporting. For P13, seeing people complaining on Reddit or Discord about being banned actually \emph{``ups my trust in the moderation system a bit more''}; for P14, expectancy was shaped through observed improvements in the broader gaming community:

\begin{quote}
    \emph{``[Counter Strike] was an absolute cesspool back in the early-mid 200s... things have gotten so much better''} - P14
\end{quote}

On the other hand, P18 felt that \emph{``nothing's really changed''}, which led them to not believe in the reporting system \emph{``it genuinely doesn't matter... I feel like it just doesn't work''}. Such experiential factors shape beliefs on how attainable positive outcomes are, which are core to expectancy. 

\subsubsection{Belief is Affected by Normative Alignment}

Participants' belief that reporting had value was driven by how well their own mental model aligned with both the reporting system and other players. Misalignment in the (subjective) definitions of punishable toxicity and the consequences of submitting a report could create inconsistent perceptions of enforcement, causing players to fail to expect the system to work as they believed. 

\paragraph{Alignment on Reporting Purpose}

Even though there are shared outcomes that players expect, player attitudes and perceptions towards how their actions lead to the outcome were varied. Some participants had a stronger view of their report as a direct signal of wrongdoing, where it logically follows that wrongdoing should be punished, and if nothing happens, then the system is broken. In such cases, reporting represents a near-causal mechanism --- you report, and the system acts: 


\begin{quote}
    \emph{``If that person is not punished ...maybe there's something wrong with the system''} - P3
\end{quote}

On the other hand, others had a more relaxed attitude towards reporting in that it acts as a single vote for toxicity within the broader scheme:

\begin{quote}
    \emph{``It's creating a paper trail, it's creating an inaudible event that even if you know this person gets reported and nothing happens this one time, if other people run into toxic behaviours in the future from this person, it's more likely to get picked up on} - P14
\end{quote}


These participants accepted that usual implementations of reporting systems mean that not everyone who behaves badly in their eyes will get punished, even though this acceptance can be begrudging at times --- \emph{``it's just like that's just how the system is''} (P9) and \emph{``I kind of learned to not expect anything from it''} (P13). In these cases, reporting is a more probabilistic indication of wrongdoing. Overall, individual attitudes toward reporting purpose shape people's expectations as to whether reporting can meet their desired outcomes. 

\paragraph{Alignment with Other Players}

All participants believed the likelihood of punishment to be dependent on the number of reports levied on a perceived toxic player, where actions were more likely to be taken when multiple players reported someone, rather than just a single individual. As such, some players perceived the process of addressing and resolving the reported behaviour as a community effort rather than a purely individual one:


\begin{quote}
    \textit{``It's not just about your individual report. It's more like a community-driven process, where members collectively vote on whether a person should be punished for toxic behaviour''} - P8
\end{quote}


Thus, it is through the collective effort of multiple reports that the system can determine whether a player is consistently toxic or only exhibits toxic behaviour occasionally. This was perceived to help ensure fairness, particularly in the anonymous environment of online games:

\begin{quote}
    \textit{``For some kind of fairness in the system, I think it makes sense for games to rely on a huge pool of complaints. So like a much larger pool than in real life" } - P1
\end{quote}

Yet, not only does this misalign with the perception of reporting as a near-causal process, but it also means that an individual's belief in the reporting system to attain value is mediated by trust in the community to use the reporting system properly --- as P17 stated \emph{``If nobody reports, then nothing is gonna happen. Nothing is gonna change... but we can't force people''}. Yet several participants had low faith in other players to use the system properly. Even in a world where players who report are well-meaning, the community representation towards collective judgment on toxic behaviour is influenced by individual norms:

\begin{quote}
    \textit{``Individuals perceive things differently in every aspect of life, including games. What feels toxic to me might not feel toxic to another"} - P3
\end{quote}

Thus, while collective judgment has the benefit of helping mitigate abuse of the reporting system, the lack of a clear shared vision for players to judge toxicity and rule-breaking means that collective judgment can misalign with any individual's own reporting model and definitions of toxicity --- \emph{``it's kind of a loose system, it's a very emotion-based system''} (P2), which can cause the perceived outcomes to be inconsistent with individual expectations.

\subsubsection{Belief is Affected by Feedback and Transparency of Outcomes}

Finally, participants' belief that reporting could lead to desired outcomes was affected by the transparency of the entire reporting-to-punishment process, and relatedly, whether feedback was provided after their reporting action. With more opaque processes, participants exercised a degree of guardedness that could lead them to question the efficacy of the system and whether their report even matters, eroding expectancy. 

\begin{table*}
\centering
\caption{Agreement levels to the questions on reporting feedback from 1 (Strongly disagree) to 5 (Strongly agree) [n=98]}
\renewcommand{\arraystretch}{1.5}
{\rowcolors{2}{white}{mypink2} 
\begin{tabular}{c C{7.5cm} c c}

\hline
\textbf{Question ID} & \textbf{Statement} & \textbf{Mean} & \textbf{Std. Dev} \\ 
\hline
    \noalign{\vskip 1mm}   
    Q10 & It is important for me to know whether the player I have reported has faced punishment & 3.69 & 1.05 \\
    Q11 & It is important for me to know the player I have reported has evaded punishment & 3.48 & 1.11 \\
    Q12 & If they have evaded punishment despite my report, I feel like I would want to know why & 3.79 & 1.15 \\
    \noalign{\vskip 1mm}   
\hline

\end{tabular}}
\Description{Table of questions related to receiving feedback of the punishment result after reporting, along with their mean and standard deviations along a 5-point Likert Scale}
\label{table:feedback}
\end{table*}

\paragraph{Presence of Reporting Feedback}

When reporting systems consistently offered meaningful and timely feedback, players were more likely to believe in the process and continue reporting toxic behaviour. Such feedback made them feel like their efforts contributed to a healthier community, reinforcing that desired outcomes were being met:

\begin{quote}
    \textit{``Sometimes you do get feedback that a player you reported recently was punished. I guess that’s like, a lot of people do find that satisfying''} - P5
\end{quote}


The quantitative survey results in Table \ref{table:feedback} also provided support for this, as respondents indicated desires to receive feedback. Conversely, a lack of feedback eroded belief and discouraged players from engaging with the reporting process --- if a player did not get feedback, they may think their report was being ignored, e.g. \textit{``if I report somebody and I don’t get any notification, I feel like they don’t take it seriously"} (P18). How feedback was presented mattered as well:

\begin{quote}
    \textit{``It would be nice to know the actual punishment that was given to the person. Right now, you just get a message, but it doesn’t tell you what the thing was”} - P2
\end{quote}

This missing information raised doubt within this participant about whether their reports were truly being processed or if the feedback was merely superficial, as P2 continued: \textit{``I feel like it’s a very placebo type of thing that they put in”}. 

\paragraph{Transparency of the Entire Reporting and Punishment Process}

The desire for feedback emerged directly from the lack of transparency often associated with reporting systems, including ambiguity around how reports are handled, the criteria for determining punishments, and inconsistencies in the outcomes of reported cases, which could reduce trust that their reporting matters and mediate the expectancy of their reports. P4 expressed that this lack of transparency led them to suspect that reporting could be influenced by personal bias. They shared that, \textit{``the decision-making process can be unclear. Punishments might vary, which is why I don’t fully trust the reporting system”} (P4). Similarly, P10 cynically suggested that this opacity and vagueness might be an intentional design by the developers:

\begin{quote}
    \textit{``I definitely think it’s subjective, and they kind of make it vague on purpose so that it’s easier for them to pick and choose. This way, they can defend whatever decisions they make because the rules are intentionally unclear}” - P10
\end{quote}

The lack of transparency could also tie into the heavy automation present in many moderation systems for large multiplayer games\footnote{e.g. \url{https://support-leagueoflegends.riotgames.com/hc/en-us/articles/201752884-Player-Reporting-Guide-and-FAQ}}. Players held varying levels of trust in reports processed by an AI versus by a human, which would impact how they expected their report to be handled. Overall, the general trend was an increased trust in human moderation, as players believed that fellow humans were better equipped to handle complex cases and ambiguous situations requiring emotional judgment:

\begin{quote}
    \textit{``The advantage of humans handling the report system is their ability to understand the context of a situation, including factors like game state, player intent, and chat history} - P4
\end{quote}


However, some players argued that AI may perform better than humans in certain scenarios because, \textit{``every human can make errors”} (P3). Players also acknowledged AI’s ability to handle reports on a large scale efficiently, as P18 highlighted, \textit{``Every day, I’m sure there are more than 100,000 reports. Many people are gaming, and it’d be difficult for a team of human staff to review every single one of them”}. Some players appreciated that AI systems can act as an early warning mechanism for inappropriate behaviour. P15 stated, \textit{``if bad content or a violation of the community guidelines is coming from your account, you should receive a warning for it.”}

\subsection{\emph{RQ4: How do the prior factors play a role in broader perceptions of trust and expectations regarding the game?}}

This section considers how the factors that affect player expectancy translate into overarching attitudes about the game but its surrounding ecosystem, including its players, the developer's reputation, and the overall experience regarding toxicity and perceived fairness. 

\subsubsection{Attitudes Towards Game Developers and the Broader Ecosystem}

While attitudes towards game developers can affect players' expectancy for outcomes, the opposite also holds --- the a priori factors that affect expectancy can also affect future trust and attitudes, namely, how players perceive the role of the developer and their reputation regarding accepting or rejecting toxic behaviour:

\begin{quote}
    \emph{``I feel like Riot [author note: Riot = Riot Games, a game developer and publisher company notable for League of Legends and Valorant] expects a level of toxicity that they deem as like coming with this type of competitive game... maybe when I play this game I should go in expecting like a baseline level of toxicity''} - P6
\end{quote}

Participants admitted the fact that game companies are likely fine with having some toxic players on their platform as well, because \emph{``taking them off their platform would lose money''} (P16) and \emph{``if there's more players ... they're gonna have income''} (P15). Yet, although players highlight these issues, their continued gameplay suggests a complex relationship between frustration with toxicity in the game, cynicism with developer motives, reporting habits, but also a desire to continue playing. 

In contrast, positive factors affecting expectancy in the reporting system can facilitate goodwill towards the game developer and encourage continued gameplay. For P19 (who plays \emph{Valorant}), the sentiment \emph{``that action is being taken... I still trust that the [reporting] process is done appropriately''} is important, as it affects \emph{``how much I would enjoy the game''}. P19 mentions that \emph{Valorant} is:

\begin{quote}
    \emph{``one of the few games where girls also play... I have girl friends who share the same sentiment, where they also feel the need to report players when they say unacceptable things''} - P19
\end{quote}

Overall, our findings here suggest that reporting and punishment systems not only impact in-game experience but also play a significant role in shaping the broader ecosystem around specific games, including highlighting the motives of the developer and promoting positive outcomes such as diversity and inclusivity. 

\subsubsection{Normalization and Acceptance of Toxic Behaviour}

Perspectives on how reporting and moderation are handled led to questions on whether toxic behaviour can ever be truly addressed in games. From a more cynical approach, P7 indicated that perceived ineffectiveness of the reporting system normalizes toxicity, because \emph{``everyone's coming on to play a game to sort of chill and blow off some steam... maybe just being toxic or something helps them''} while knowing they can get away with it. Many participants indicated how toxicity was nowadays inherently ingrained in certain games:

\begin{quote}
    \emph{``[League of Legends] is a game that's so old that people kind of know what to expect from it''} - P2
\end{quote}

\begin{quote}
    \emph{``I feel like [toxicity]'s just kind of expected, especially with let's say Call of Duty''} - P9
\end{quote}

With an ingrained reputation and the perceived ineffectiveness of the primary tool players can use to combat toxicity, players indicated a sense of helplessness --- \emph{``it's kind of impossible [to change] because just like how gaming in general inherently is''} (P15) --- even when reporting is not the sole way of addressing toxicity. Despite this, when people give up on thinking that the community can ever improve, they lose one of the two motivators for reporting. Some participants felt discouraged about their actions when they felt nothing was changing. Yet when reporting remains the primary in-game option to combat toxicity, this leads to a potentially negative feedback loop. 

On the other hand, not everyone felt hopeless, with some participants indicating that the gaming community was getting better. These observations helped participants believe that things were changing positively, and partially played a role in their continued reporting behaviours even if community improvement was not directly solely correlated to reporting behaviours, such as:

\begin{quote}
    \emph{``going back to like Xbox Call of Duty lobbies ... those were just a cesspool compared to today; there definitely are stronger moderation standards''} - P14
\end{quote}


Thus, if players perceived the community as getting better (whether due to reporting practices or not), then their altruistic motivations for reporting are met and they may be more motivated to continue to report (e.g. \emph{``I feel like I'm reporting in that game, it makes the environment better for other players as well. So that gives me more motivation to report in that game''} --- P19), leading to a positive feedback loop in terms of change. 

\section{Discussion}

Our findings highlight that, under an EVT perspective, reporting holds two somewhat contrasting sources of value, one driven through retributive motivations to punish the toxic perpetrator, and one driven through altruistic motivations to improve the gaming community. However, participants noted that reporting neither consistently nor effectively meets each of these outcomes, because not every instance of reporting leads to punishment, and punishment is not always conducive to the change required for broader improvement. 

We highlight that the belief that such outcomes will be reached (i.e., expectancy) is shaped by various factors such as trust based on prior experiences, normative alignment with the community, and system transparency. Importantly, belief in reporting contexts is shaped, not by self-efficacy like prior EVT studies \cite{shinWhenMotivationIsnt2025, rachmatullahModelingSecondaryStudents2021}, but by surrounding sociotechnical factors. Participants sometimes indicated low expectancy when they felt like the developer did not care, or the reporting system was too opaque, even when reporting was valued. In our discussion, we expand on two problems related to each of the EVT constructs. For expectancy, we consider ways for developers to increase user belief in the system and their reporting actions; for value, we consider ways to align two contrasting motivations that may sometimes have misaligned consequences. 

\subsection{Bridging the Gap of Belief --- Design and Challenges Regarding Expectancy}

Various factors that mediate people's \textbf{beliefs} affect perceptions of reporting's effectiveness. Much prior past work highlights the doubt that players can hold towards perceived ineffective reporting systems \cite{kouFlagFlaggabilityAutomated2021a, pohjanenReportPleaseSurvey2018}. As with our work, players ascribe part of that fault to the developers of the game, and developers play a role in controlling these behaviours \cite{lapollaTacklingToxicityIdentifying2020, hilbersEvaluationEffectivenessSmites2015}. Our work highlights how various dimensions, such as transparency, feedback provision, and prior experiences, all play a role in establishing people's trust in the black-box reporting system and, thus, expectancy. 

In Section 2.2, we reviewed how laws that align with shared values encourage adherence, but a lack of transparency in the legal system can influence individuals' likelihood of breaking laws \cite{tyler2001obeying}. Analogizing to a gaming context, players emphasized the need for greater transparency in the reporting system and called for developers to prioritize and expedite the handling of reports. A lack of transparency and consistency in regulation lowered their trust in the system, as players could seemingly get away with infractions. Beyond regulation transparency, the provision of feedback was noted as helpful for people to know that their reports are being heard, as well as to build player trust in the system; agreeing with prior work \cite{kordyakaCuringToxicityDeveloping2021}. Given the importance of transparency and feedback, we propose that existing game reporting systems could better align with the existing Santa Clara Principles \cite{santa_clara_principles} by being more open about the moderation process and providing detailed explanations to both the reporter and the punished player.

\subsubsection{Moderation Trust with Automation}

Developers often face challenges in monitoring games and handling reports due to the large number of players. While Hilbers suggests incorporating automation to guide player behaviour \cite{hilbersEvaluationEffectivenessSmites2015}, we find that players perceived automation as struggling to comprehend complex contexts and deliver fair decisions. Human moderators, on the other hand, were perceived to provide deeper contextual understanding but were limited by resource constraints. Human judgment and bias are tied to player concerns about the inconsistency of reporting outcomes over time, emphasizing the need for a more robust system. This generally aligns with Sparrow et al.'s prior research, which highlights the many roles and mental models of AI moderation in multiplayer games --- as an unreliable police force, an unscrupulous governor, an uncaring judge, and an untiring assistant \cite{sparrowEthicalAIModeration2024a}.

The problem, common to both Sparrow et al.'s work and our findings, is that players have inconsistent mental models when faced with a black-box moderation system. Yet, we note that while AI moderation adds to this opacity, reporting systems would likely still have a level of higher judgment invisible to the users; even crowdsourced judgments had concerns over transparency \cite{kouManagingDisruptiveBehavior2017}. When faced with an uncertain system and uncertain consequences, players inevitably face challenges in trust \cite{steedmanComplexEcologiesTrust2020}. Given that trust then delegates to what players \emph{can} observe, we propose rethinking and redesigning human-AI systems for moderation \cite{lai2022human} with a focus on communication, transparency, and player perception. We consider systems that might leverage AI to support human decision-making at scale, combining the efficiency of AI with the nuanced judgment of human moderators to enhance fairness and consistency in the reporting process \cite{wang2022ml, lykouris2024learning}.  

We highlight ways that AI moderation can be designed to be more acceptable and trustworthy by players, with the assumption of benevolent intention to increase trust \cite{liaoDesigningResponsibleTrust2022}. To provide feedback desired by reporters, we envision an AI-supported triage system. AI could escalate the highest priority reports \cite{montgomeryCustomerSupportTicket2018} or cluster similar reports \cite{royClusteringLabelingIT2016}. Overall, to extend transparency in driving trust, reporters could view the status of their report at all times, from when it is in this AI-supported queue, to when it is being reviewed by humans or AI, to whether punishment is levied or not and why, and perhaps even afterwards, whether the offending player has actually reformed. Under this perspective, AI would serve as a proxy for information that is presently missing for reporters. 

Instead of optimizing and presenting the AI (and moderation as a whole) as an authoritarian source of truth, developers should highlight and communicate how they are using the AI in their moderation pipeline, provide examples of AI judgment, and provide documentation and statistics for transparency purposes \cite{liaoDesigningResponsibleTrust2022}. To garner trust, AI-assisted processes should acknowledge that AI can make mistakes \cite{zerilliHowTransparencyModulates2022}; in such cases, perhaps opportunities to appeal are warranted \cite{sparrowEthicalAIModeration2024a}. In general, trustworthiness in this context is fostered through participants forming a mental model in which they understand, are confident in, respect, and are comfortable with automation \cite{schaeferMetaAnalysisFactorsInfluencing2016}. If developers are to use automation as part of their report-handling, then ultimately trust, and thus user expectancy, can only be built through understanding the system rather than fearing it. 

\subsubsection{Trade-offs for Transparency}

Yet, while transparency is vital for trust \cite{liaoDesigningResponsibleTrust2022, zerilliHowTransparencyModulates2022, bhatiaBuildingTrustTransparency2024}, we acknowledge that complete transparency might not be appropriate. Specifically, revealing the full details of how the reporting system detects and reviews toxic behaviour could enable toxic players to manipulate the system and avoid detection. This issue parallels the broader debate about the extent to which laws should list reasons for crimes. Some scholars argue that listing reasons helps clarify legal boundaries \cite{hart2012concept} and reduces misuse of the law \cite{fuller1969morality}, thereby preventing unintentional violations and protecting citizens' rights. However, others caution that overly detailed reasoning can introduce significant drawbacks such as increased legal complexity \cite{schauer1991playing} and create exploitable loopholes \cite{posnereconomic, sunstein2018legal}. In a reporting context, if users know a word would definitely be flagged by the rules, then they may be able to bypass it by using a variant (e.g. by changing a letter). Thus, a potential compromise is adopting principle-based provisions, where general guidelines are established, and specific applications are clarified through judicial interpretations or guiding cases \cite{alexy2010theory, llewellyn2012bramble}. 

Presently, community guidelines are usually provided upfront \emph{once} but not reviewed as an ongoing activity. We propose weaving community guidelines into the constant gameplay experience where possible, for instance, rotating important guidelines on waiting screens or log-in pages with specific examples or interpretations, or perhaps even incorporating them during periods of low-action into the gameplay loop (e.g. as reminders in the messaging system). This would serve several purposes in improving the expectancy of reporting outcome --- highlight that toxicity is a problem that is taken seriously within the game, raise ongoing community awareness of toxicity (and that reporting is a tool to combat it), and reinforce community standards on toxicity that can otherwise be subjective. We also highlight that there are ways for developers to demonstrate transparency beyond revealing details of the process, such as providing broader metrics, diagnoses, stats, and documentation regarding moderation \cite{liaoDesigningResponsibleTrust2022}. Providing such broader, anonymized reports may help with transparency and accountability without the ethical concerns (of revealing details to be taken advantage of) or privacy concerns (of naming and shaming, which could lead to witch hunts \cite{bergstromDontFeedTroll2011}). 

\subsection{Building a Better Community - Understanding and Aligning Reporting Values}

Our findings suggest that reporting was expected to lead to punishment, and while punishment was regarded as an effective way to reduce toxic behaviour (although possibly bypassed through new accounts), reporting was not always perceived to be effective in reaching the punishment stage \cite{pohjanenReportPleaseSurvey2018, kouFlagFlaggabilityAutomated2021a}. Kou also found that reporting is used for reasons outside of dealing with toxicity, such as its use in coordinated, targeted contexts or its use in venting frustration, dissociating reporting from actual ``toxicity'' \cite{kouFlagFlaggabilityAutomated2021a}. We find that reporting is used for two sometimes conflicting motivations --- as an idealized way to combat toxicity, and as a self-fulfilling way to punish people whom you have had a poor experience with. Yet, despite our participants' distrust of reporting systems that agree with prior research, almost all participants continued to use it (also refer to Q9 in Table \ref{table:expectancies}) in hopes of a better community, representing a possibly Sisyphean hope that reporting can be used in its ideal context as leading to a better community. 

\subsubsection{Moving from Individual Punishment to Broader Community Improvement}

Beres et al. suggested that an impediment to reporting is the normalization of toxicity --- people lack the motivation to report when toxic behaviours are rationalized within the community context or when they distance themselves from the other players \cite{beresDontYouKnow2021}. The latter concept of ``distancing'' was also reinforced through Kou's consideration of permanent bans \cite{kouPunishmentItsDiscontents2021}. Whereas permanent bans are used as a disciplinary method to induce change, Kou highlights how their deeper functions are to produce a stereotype of ``the most toxic'' player in the community \cite{kouPunishmentItsDiscontents2021}. These prior works highlight the deep division of the community into ``toxic'' and ``non-toxic'' players as two groups in tension; this division both motivates and undermines the act of reporting. On one hand, it supports reporting as a way to exclude toxic players (the retributive motivation); on the other, it fosters skepticism about its effectiveness, as the community may not always improve (the altruistic motivation). Poeller and Robinson highlight how reporting fails to address the underlying problem regarding toxic communities \cite{poellerMuteBlockPunish2024}.
The fracture in the community contrasts with more recently suggested restorative justice approaches to moderation. Restorative justice defines infractions as a violation of people and punishment, rather than a violation of the rules; justice comes from repair and reconciliation rather than punishment \cite{umbreitCrimeVictimsSeeking1989, clarkThreeRsRetributive2008, wenzelRetributiveRestorativeJustice2008}. Kannabiran espouses the importance of forgiveness when human relationships are transgressed, with repair coming from a chance to make amends and admit mistakes \cite{kannabiranAmSorry2021}. Vasalou et al. implemented a system for repairing trust based on apology and forgiveness, finding that these factors allowed victims to recover trust \cite{vasalouPraiseForgivenessWays2008a}.

Restorative justice and forgiveness are promising directions for the altruistic goals of reporting. Yet there are challenges, as we highlight three interrelated open questions. Firstly, Xiao et al. indicated that restorative justice is a community effort, yet victims, offenders, and moderators in gaming communities may lack a shared sense of community \cite{xiaoAddressingInterpersonalHarm2023}; --- thus, beyond reporting, \emph{how can we reinforce the shared identity of players to re-form a community as a whole, especially given the ephemeral context of team-based multiplayer games}? Secondly, Kannabiran brought up challenges with restorative justice and forgiveness as a whole \cite{kannabiranAmSorry2021} --- thus, \emph{how do we ensure that forgiveness takes place after reporting in a manner that is adequate, appropriate, and fair to the victim}? Lastly, Reid et al. explored novel ways for people to deal with toxicity, but also found that people found that players were skeptical of tools that focus on support rather than discipline \cite{reidFeelingGoodControl2022}. People desire discipline as a form of revenge against toxicity --- thus, \emph{how do we validate the negative experience and deal with the frustration and spite that a reporting player might have held towards the other}? 

\subsubsection{Finding Common Values}

These challenges converge on the issue of \textbf{aligning the sometimes divergent reporting motivations towards a unified goal}. Punishment and restorative justice are not inherently oppositional, but punishment must be \emph{meaningful} to induce change. Our challenges double as design suggestions. For instance, we consider solutions such as fostering a shared community identity through value alignment. Doing so fundamentally may require re-thinking design from the ground up, to reconcile opposing values between reporting players and reported players, and even within groups of reporting players. Design methodologies such as Value-Sensitive Design \cite{oosterlakenApplyingValueSensitive2015} and Ally-Opponent Understanding \cite{jangAllyOpponentUnderstandingCoexistence2019} provide promising baselines for building with multiple groups of values. Both these approaches espouse the importance of talking to stakeholders early, understanding their perspectives, and evaluating by multiple groups of stakeholders. 

Although we have not engaged in such a design exercise, we provide some initial brainstormed examples. Games could provide non-identifying glimpses and light tasks about the other players to humanize them and support camaraderie in initial team-building (e.g. ``this player has not played in several months --- help them adapt!''). In doing so, we point towards a shared goal that can be accomplished collaboratively, even in a competitive environment. We also encourage restorative interactions, such as an apology by the toxic player, and empowering the victim of toxic behaviour to articulate the impact of toxicity and involve them in the resolution. These steps may help bridge the gap between punitive and altruistic approaches, advancing toward a better community.

\section{Limitations and Future Work}

Limitations are mainly tied to the participants and their engagement in the study. Some of these limitations are fundamental to user studies, for example, self-selection bias \cite{elstonParticipationBiasSelfselection2021} --- people who are likely to want to contribute to the study may already have some vested opinions in reporting systems compared to those who do not, which may skew the presented perspectives. As our research focuses specifically on reporting practices, social desirability response bias may also exist --- people may try to respond to paint themselves in a morally correct light \cite{randallSocialDesirabilityResponse1991}. In this study, we hypothesize that such effects may cause people to highlight the altruistic motivation for reporting and minimize the retributive motivations. This could be exacerbated by the fact that our study focused on people's recounting of their experiences, which may be subject to recall bias due to selective memory \cite{shieldsRecallBiasUnderstanding2010}. Reporting could take place as singular moments far in the past, meaning that recollective feelings may have changed or become muted over time. To alleviate these problems, future work can incorporate a general sample of gameplay to verify whether the recounted data holds during actual observed experiences across a wider sample of players. 

We also highlight that our work is still exploratory in understanding people's reporting systems and their relationship to EVT. While our qualitative insights provide an initial understanding of our participants' values and expectancies, future work can perform a confirmatory survey with a large sample of participants using existing scales for motivation and self-efficacy (e.g. as in \cite{shinWhenMotivationIsnt2025, yangInvestigationCollegeStudents2013}), to see if similar results hold through further triangulation \cite{carterUseTriangulationQualitative2014a}.

Extending on the sample of interview participants, most were young. Although this was somewhat expected given the domain and samples seen in prior research, participants did mention that age can affect players' toxic behaviour and their engagement with the reporting system, for example, several participants mentioned that they had been more likely to abuse the report button when they were younger. Most participants also discussed competitive or team-based games; future work could explore reporting experiences in co-operative games. Finally, the general opacity of different reporting systems makes it hard to make concrete judgments; many perspectives in this work rely on participant anecdotes, experiences, and their best guesses.

\section{Conclusion}
We holistically investigated reporting systems in multiplayer online video games through expectancy-value theory. Using a survey and follow-up interviews, we found that players' values in reporting are both altruistic and retributive, viewing reporting as a means of community betterment and personal venting, respectively. However, players often perceive that their desired outcomes expected from these motivations are not always fully realized, with the degree to which players believe in these outcomes influenced by factors such as the players' personal attitudes, their trust in the community and the developer, and design factors of the reporting system including its level of transparency, the provision of feedback, and the degree of automation. These perceptions affect players' beliefs and attitudes, shaping their view of developer reputation and potentially leading to the normalization or passive acceptance of toxic behaviour in games. Based on our findings, we propose guidelines for designing reporting systems and fostering healthier gaming communities.

\begin{acks}

\end{acks}

\bibliographystyle{ACM-Reference-Format}
\bibliography{sample-base}


\end{document}